\documentclass[runningheads]{llncs}
\usepackage[T1]{fontenc}
\usepackage{graphicx}
\usepackage{adjustbox}
\usepackage{caption}
\usepackage{subcaption}
\usepackage{amsmath}
\usepackage{amssymb}
\usepackage{times}
\usepackage{epsfig}
\usepackage{multirow}
\usepackage{hyperref}
\usepackage{color}
\usepackage{graphicx}
\begin{document}
\title{Virtual-Reality based Vestibular Ocular Motor Screening for Concussion Detection using Machine-Learning}
%
%
\titlerunning{VR-based Screening for Concussion using Machine-learning}

%

\author{Khondker Fariha Hossain \inst{1} \and Sharif Amit Kamran \inst{2} \and Prithul Sarker \inst{3} \and Philip Pavilionis \inst{4} \and Isayas Adhanom \inst{5} \and Nicholas Murray \inst{6} 
  Alireza Tavakkoli \inst{7}}
\authorrunning{K.F. Hossain, et al.}
 %
\institute{University of Nevada, Reno, NV 89557 \\ 
\email{khondkerfarihah@nevada.unr.edu \inst{1}}, 
\email{skamran@nevada.unr.edu \inst{2}},
\email{prithulsarker@nevada.unr.edu \inst{3}},
\email{ppavilionis@unr.edu \inst{4}},
\email{iadhanom@nevada.unr.edu \inst{5}},
\email{nicholasmurray@unr.edu \inst{6}},
\email{tavakkol@unr.edu \inst{7}}
}

%
\maketitle              
\begin{abstract}
Sport-related concussion (SRC) depends on sensory information from visual, vestibular, and somatosensory systems. At the same time, the current clinical administration of Vestibular/Ocular Motor Screening (VOMS) is subjective and deviates among administrators. Therefore, for the assessment and management of concussion detection, standardization is required to lower the risk of injury and increase the validation among clinicians. With the advancement of technology, virtual reality (VR) can be utilized to advance the standardization of the VOMS, increasing the accuracy of testing administration and decreasing overall false positive rates. In this paper, we experimented with multiple machine learning methods to detect SRC on VR-generated data using VOMS. In our observation, the data generated from VR for smooth pursuit (SP) and the Visual Motion Sensitivity (VMS) tests are highly reliable for concussion detection. Furthermore, we train and evaluate these models, both qualitatively and quantitatively. Our findings show these models can reach high true-positive-rates of around 99.9 percent of symptom provocation on the VR stimuli-based VOMS vs. current clinical manual VOMS.

\end{abstract}

\section{Introduction}

 In the past decade, there has been an increasing amount of awareness of sports-related concussions (SRC) associated with physical activity and athletics. The American Medical Society for Sports Medicine defines concussion as a traumatically affected temporary disruption of brain function implicating complex pathophysiological processes \cite{harmon2013american}. The long-term impact of SRC is only now being examined, and its effects on athletic performance and the long-term health of athletes. Regardless, concussion assessment remains controversial for diagnosis, management, and return-to-play guidelines for sports-related concussions. A prior study has revealed that sport-related concussion affects up to $38$ million people annually in the United States alone. At the same time, up to $50\%$ of these injuries are unreported as athletes recognize the symptoms but do not want to miss playing time due to the lengthy nature of SRC recovery\cite{centers2007nonfatal,langlois2006epidemiology,centers2015report}.

The Vestibular Ocular-Motor Screen (VOMS) is a standard SRC assessment tool designed to assess symptoms associated with the brain's vestibular and ocular motor systems. Developed at the University of Pittsburgh Medical Center, the VOMS involves a battery of tests that challenge the subject’s eye tracking specific to smooth pursuit, saccades (horizontal and vertical), and near point convergence (NPC). It also challenges the vestibular system with the VOR and vestibular motion sensitivity (VMS) tests.  The subject is asked to rate symptoms at baseline and after each test: headache, nausea, dizziness, and fogginess. A 10-point Likert scale is used to rate the symptom with a zero, meaning the symptom is not present to a 10 in an emergency medical situation. Typically, the VOMS is administered clinically with the clinician using their fingers as targets for the test, or sometimes a tongue depressor with a 14-point dot at one end is used for the eye targets. A measuring tape is used for the NPC test, and some form of a metronome is used for the saccades, VOR, and VMS. The clinician should be 3 feet from the subject when holding the targets in the specified position for each test. For example, the smooth pursuit test involves moving the target a total of 3 feet horizontally, then vertically. The targets for the saccades need to be 3 feet apart. Because of the varying arm length of each clinician and their perception of the 3-foot distance, this can vary among clinicians.  Applying the VOMS in a virtual reality environment would provide a means of standardization among clinicians \cite{winkler2016adult}.

Recent studies found that shifting visual input, vestibular and proprioceptive information are some features of the Egomotion phenomenon, which has been broadly studied in an experiment called  "moving room paradigm" \cite{lestienne1977postural,lee1974visual,lee1975visual,stoffregen1998postural}.  Egomotion defines as body motion or visual scene motion response to optic flow \cite{warren1976perception}. So, contemporary analysis of optic flow with different visual stimulus patterns can assist in comprehending the influence of the visual field motion in subjects suffering from a concussion or mild traumatic brain injury(mTBI) \cite{beer2002areas}.Virtual reality (VR) is one part of Extended reality (XR) which also contains augmented reality (AR) and mixed reality (MR). VR can be depicted as computer-generated simulated environments in real or imagined worlds where the user can interact with the virtual environment without utilizing visual cues from the real world \cite{howard2017meta,chuah2018and}. VR technologies project the user into a three-dimensional (3D) environment where the user enters the 3D world by wearing a head-mounted display (HMD). A detection system integrated into the HMD senses the user's physical motion, which is transformed into computerized data \cite{nolin2012virtual}.

The manual VOMS depends on the subjective reporting of provoked symptoms, so it is vital to comprehend the eye movements associated with them. In our experiment, we operated a virtual reality (VR) headset to produce a VOMS assessment to eliminate subjectiveness and advance the objectiveness of the assessment. We experimented on the eye movement data of Smooth Pursuit and Visual Motion Sensitivity to detect concussion based on the prominent features. To validate our results, we compared them with manual VOMS scores (0-10) for the reliability check.

\section{Related Work}
The Standard Assessment of Concussion (SAC), created in 1997, includes orientation, focal deficit screening, exertional exercises, concentration assessments, and immediate and delayed verbal memory tests. The SAC evaluates mental health in four areas: orientation, short-term memory, concentration, and delayed recall. Even though the SAC is a component of the initial concussion evaluation, it is insufficient for a thorough concussion evaluation. The SAC score alone cannot be used to identify concussion severity or encourage a player's return to play.

A highly integrated neural network made up of cortical, subcortical, and brain stem regions, the spinal cord, and peripheral nerves are responsible for the human ability to maintain balance. Guskiewicz et al. created the Balance Error Scoring System (BESS) to evaluate an athlete's balance following a concussion \cite{guskiewicz2001postural}. The athlete must complete trials for the BESS test. The BESS score is based on how many mistakes were made on each trial.

SCAT-3 is the third version of the sports concussion assessment tool (SCAT) after SCAT-2 received concerns regarding the sensitivity and specificity of the results. In addition, SCAT-3 includes a neck/cervical spine examination that was not previously part of another SCAT testing. Therefore, athletes 13 years and older are advised to take the SCAT-3 test.
The nature of the tests described above is entirely subjective, and the tests take much time. The K-D test is a quick number naming exam that may be completed in under one or two minutes. It consists of a demonstration card and three subsequent tests. The K-D test has been used to assess a variety of illnesses, including visual impairments. However, even without a concussion, scores on the K-D test typically rise over time as a result of learning effects brought on by repeated testing. The test primarily compares variations between baseline and post-injury assessments.

VR is becoming more popular for traumatic brain injury (TBI) evaluation and recovery. The results from various literature imply that VR may be a more sensitive technique for SRC assessment despite the small sample size and constrained scope. A VR environment may be more ecologically valid than a conventional laboratory setting for detecting concussions. The stimuli in VR are multisensory and accurately simulate physical experiences. Using VR technology, Teel et al.\cite{teel2015validation} applied a therapeutic variation of the "moving room" paradigm. The results show that the overall accuracy with a particular threshold requires further evaluation. Wright et al. \cite{wright2017visual} also noted that SRC employing the virtual environment TBI screen results in a destabilizing effect of visual field motion.

The VOR was evaluated by Murray et al. \cite{murray2014assessment} using the Wii Fit Soccer Heading. The eye movements show that although there were no group differences, the concussion group exhibited more gaze deviations from the center than the control group. Only a few research have used VR techniques for SRC assessment and therapy. However, neuropsychological testing and VR postural control indicate a positive advancement in SRC management.

\section{Methodology}
In this section, we describe the supervised methods we have utilized for classifying the concussed vs. non-concussed signals and the initial hyper-parameters chosen for them. First, we discuss the supervised techniques utilized for quantitative evaluations in sub-sections \ref{nb}, \ref{dt}, \ref{rf}, \ref{svc}, \ref{ada}, \ref{gpc}, \ref{lr}, and \ref{perc}. In the next two subsections \ref{iso} and \ref{one-svc}, we discuss two other methods for qualitative evaluation and visualization. All of the supervised methods were trained and evaluated using scikit-learn library \cite{pedregosa2011scikit}.

\subsection{Naive Bayes}
\label{nb}
Naive Bayes (NB) classifier \cite{rish2001empirical} is the type of probability-based method that extracts important features that describes model attributes. It is a supervised learning algorithm based on applying Bayes’ theorem with the “naive” assumption of conditional independence between every pair of features given the value of the class variable. For parameter, we use the default \textit{var\_smoothing}$=10e-9$.

\subsection{Decision Tree}
\label{dt}
Decision Tree \cite{kingsford2008decision} is a non-parametric supervised classifier technique that can be utilized for both classification and regression tasks. The model predicts the value of a target variable by learning straightforward decision rules deduced from the data features. For parameters, we select splitting \textit{criterion}=\textbf{gini}, \textit{min\_samples\_split}$=2$, and \textit{min\_samples\_leaf}$=1$.

\subsection{Random Forest}
\label{rf}
Random forests classifier \cite{breiman2001random} is an ensemble method for classification, regression, and other tasks that builds multiple decision trees on various sub-samples of the dataset at training time and builds consensus based on the outputs. During test time, the predicted class is the majority selected by the trees. For our model, we select \textit{n\_estimators}$=100$, splitting \textit{criterion}=\textbf{gini}, \textit{min\_sample\_split}$=2$, \textit{max\_features}=\textbf{sqrt}.

\subsection{Support Vector Classifer}
\label{svc}
Support Vector Classifier (SVC) \cite{chang2011libsvm} is a variant of Support Vector Machine (SVM), which supports different linear and nonlinear kernels. It is a supervised technique for classification or regression tasks. Like SVM, SVC uses the hypothesis space of a linear function and has been applied to different biomedical signal processing techniques such as ECG \cite{zhao2005ecg}, EEG \cite{subasi2010eeg} and EMG \cite{toledo2019support} signal classification. For parameters, we use Radial-basis Function (RBF) as kernel and select kernel-coffeicient, $\gamma$ based on the variance of input features.

\subsection{AdaBoost}
\label{ada}
Adaboost \cite{ratsch2001soft}, which stands for Adaptive Boosting algorithm and is an ensemble method to improve performance over linear classifier. Again, the goal is to tweak the subsequent weak learners in favor of those instances misclassified by previous classifiers. We use the Decision-tree classifier for the base-estimator,  with \textit{n\_estimator}$=50$ and a learning rate of $1.0$.

\subsection{Gaussian Process Classifier}
\label{gpc}
Gaussian Processes Classifier \cite{rasmussen2010gaussian} is a non-parametric method for classification tasks and is a generalization of the Gaussian probability distribution. Inherently, the Laplace approximation is utilized for approximating the non-Gaussian posterior. For parameters, we use \textit{optimizer}=\textbf{fmin\_l\_bfgs\_b} and \textit{max\_iter\_predict}=$100$. The optimizer, Limited-memory BFGS (L-BFGS or LM-BFGS), is an optimization algorithm that approximates the Broyden–Fletcher–Goldfarb–Shanno (BFGS) algorithm utilizing a small amount of computation memory \cite{liu1989limited}.

\subsection{Logistic Regression}
\label{lr}
Logistic regression \cite{hosmer2013applied} is a supervised learning technique for binary classification. It identifies the relationship between a dependent variable and one or more independent variables and categorizes two distinct classes. For our models, we utilize the parameters, \textit{penalty}=\textbf{l2}, \textit{tol}$=1e-4$, \textit{solver}=\textbf{lbfgs} and \textit{max\_iter}$=100$.

\subsection{Perceptron}
\label{perc}
Perceptron \cite{rosenblatt1958perceptron} is a supervised learning algorithm for binary classification that multiples the input feature vector with a weight coefficient and generates an output value based on an activation function. The learning rule compares the predicted output with the known output. If it does not match, the error is propagated backward for updating the weight. The process is reiterated until the error converges to a specific value or satisfies a criterion. For parameters, we use $\alpha=0.0001$, \textit{max\_iter}=$100$, \textit{eta0}=$1$, \textit{validation\_fraction}=$0.1$ and \textit{tol}=$1e-3$.  Here, \textit{tol} stands for the stopping criterion when (loss > previous\_loss - tol).

\subsection{Isolation Forest}
\label{iso}
Isolation Forest \cite{liu2008isolation} algorithm returns the anomaly score of each data point. It works by isolating observations by randomly picking a feature and then randomly picking a split value between the highest and lowest values of the selected feature. As partitioning creates a tree-like structure, the number of splittings required to isolate a sample is identical to the distance from the root node to the ending node. Consequently, random partitioning creates shorter paths for anomalies. Hence, when a collection of random trees together produces a shorter distance for particular samples, they are highly likely to be anomalies. We can utilize this to finding samples for concussed data points by treating as anomalies.

\subsection{One Class SVM}
\label{one-svc}
One-class SVM \cite{scholkopf1999support} is a unsupervised novelty detection algorithm. Generally, SVMs are max-margin techniques that classify binary targets and cannot model probability distribution. So, the One-class SVM idea finds positive areas for high densities of samples and negative areas for small densities of points. We can extend to our problem of finding positive areas for non-concussed data points and negative areas for concussed data points.

\section{Experimental Analysis}
\subsection{Data Collection using Virtual-Reality Headset}
\subsubsection{Equipment:}
The VOMS protocol administered in VR was conducted with an HTC Vive Pro Eye Head Mounted Display (HMD) with a diagonal field of view (FOV) of 110-degree, refresh rate of 90Hz, a combined resolution of 2880×1600 pixels, six degrees of freedom (DoF) for position and orientation tracking, and adjustable inter pupillary (IPD) and focal distances. Before beginning the exam, the participant’s nose length was measured and recorded to account for HMD being mounted onto the head. The headset was powered by an Acer Predator gaming laptop with a 7th Generation Intel Core i7 Quad-Core processor with 16GB of memory and NVIDIA GeForce GTX 1070 graphics card running Windows 10. The HTC VIVE 2.0 Hand Controller was used to receive input from participants to start the stimuli moving towards them and stop the stimuli when ocular convergence was lost (image splitting into two objects) during the NPC test and to re-create the target object during the VMS component of the VOMS. The Unity3D engine version 2019.1.6 and the Unity3D VR plugin - SteamVR - version 1.7 was used to develop the VOMS stimuli in which the VOMS protocol was simulated in a VR environment. HTC’s Sranipal SDK version 1.1.0.1 was used to read eye-tracking data from the eye trackers and will be reported in another publication. 

\begin{figure}[!tp]
    \centering
    \includegraphics[width=.9\textwidth]{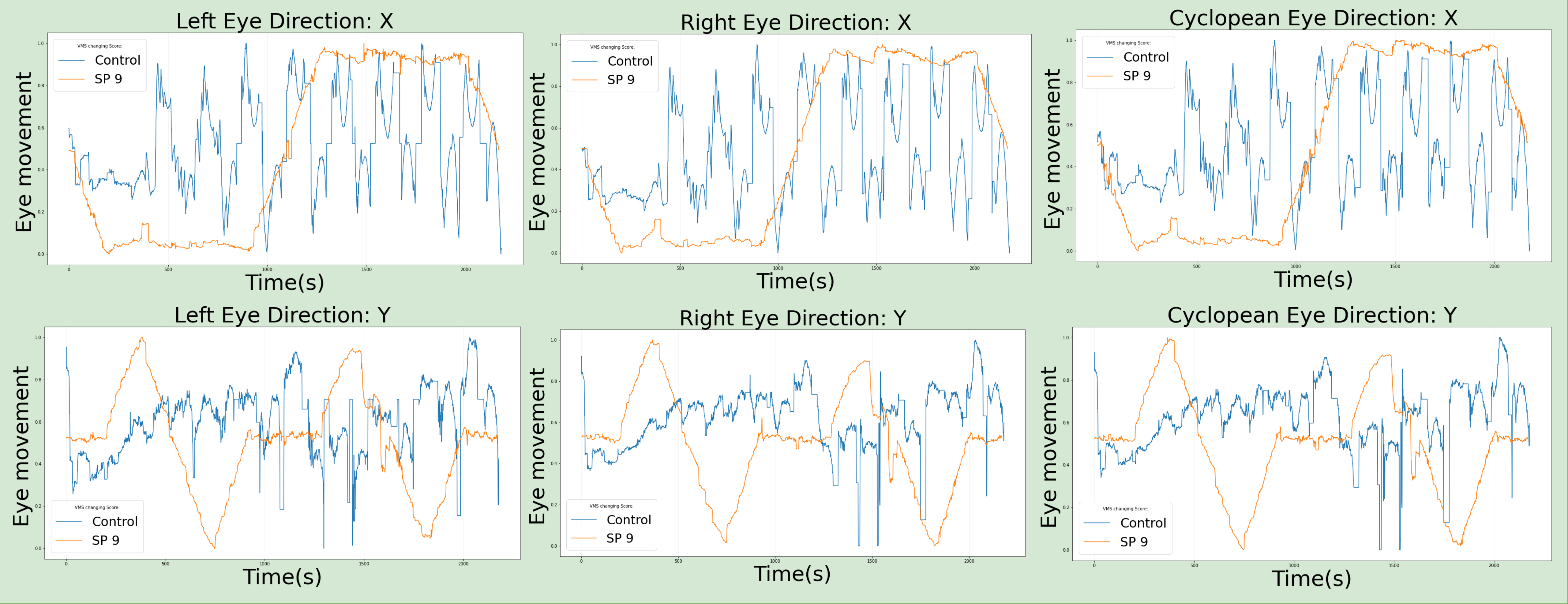}
    \caption{\textbf{Smooth Pursuit:} A Control (non-concussed) and a Concussed patient data overlapped on top of each other. The columns I and II are for x and y directions. The row I-III are for Left eye, Right eye and Cyclopean Eye direction. For each subplot, x-axis represents time and y-axis amplitude.}
    \label{fig1}
\end{figure}

\subsubsection{Procedure:}
All participants were evaluated at pre-season physicals or within 24-48 hours following the injury. All participants completed the VOMS in VR using the prior suggested guidelines. The assessment of the VOMS involves an administrator assessing baseline symptoms which consist of headache, nausea, dizziness, and fogginess.Participants were asked to state the severity of each symptom based on a scale of 0-10 before the start of the VOMS assessment and after each sub-test of the VOMS. The sub-tests were recreated in VR while following the literature standards.

\begin{figure}[!tp]
    \centering
    \includegraphics[width=0.9\textwidth]{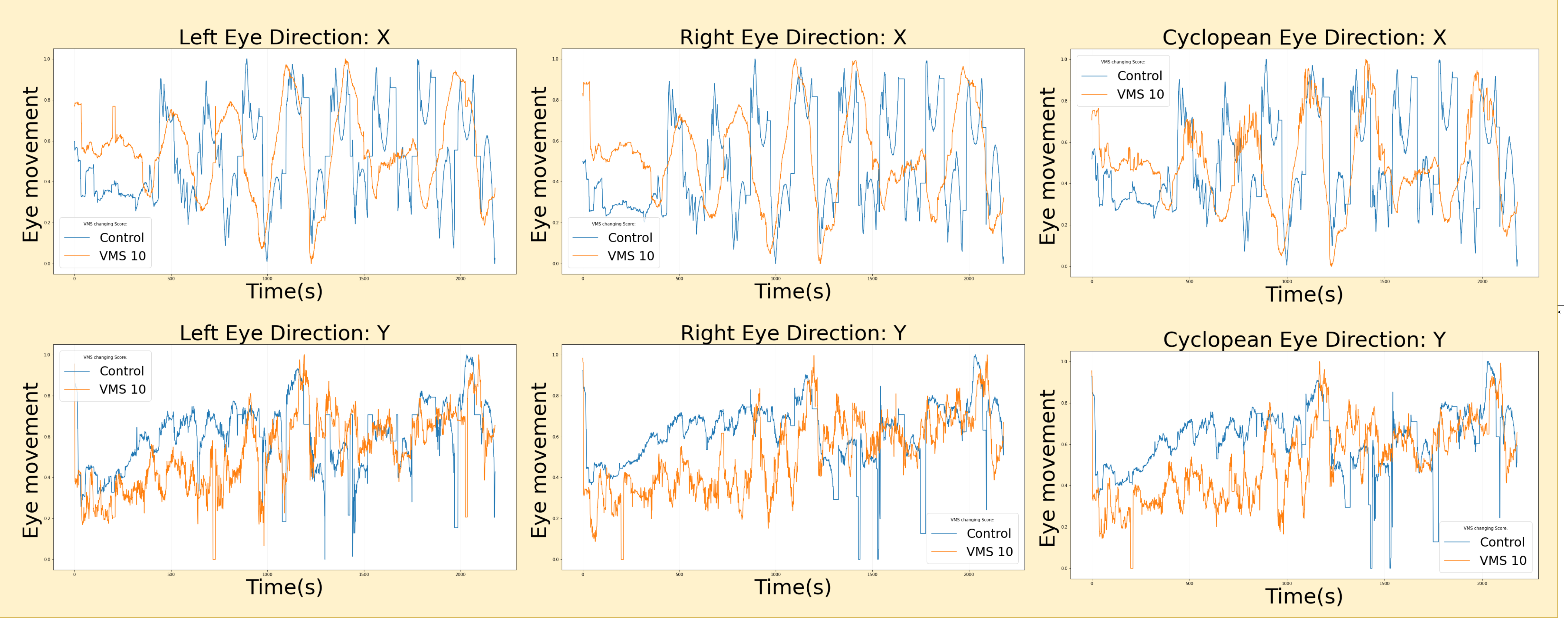}
    \caption{\textbf{Visual Motion Sensitivity:} A Control (non-concussed) and a Concussed patient-data overlapped on top of each other. The columns I and II are for x and y directions. The row I-III are for Left eye, Right eye and Cyclopean Eye direction. For each subplot, x-axis represents time and y-axis amplitude.}
    \label{fig2}
\end{figure}

\begin{itemize}
    \item \textbf{Smooth Pursuits (SP):}
    The administrator holds up the two targets and instructs the subject to keep their head still while moving their eyes from one target to the other to the beat of a metronome app set at 180 beats per minute. The targets are held three feet apart which is again estimated by the administrator.The subject was instructed to focus on the target with their eyes and follow the target while keeping their head motionless. The system would move the target in an “H” pattern. This pattern runs three feet horizontally and 3 feet vertically within the participants view.. In Fig.~\ref{fig1}, we visualize a concussed and a non-concussed patient's data for left, right and cyclopean eye direction for smooth pursuits.
    
    \item \textbf{Visual Motion Sensitivity (VMS):}
    The subject is asked to stand and is given one of the targets. They are instructed to hold the target at arm’s length at the height of their nose.  They are instructed to fixate with their eyes on the target while rotating their upper body in a 180-degree motion (from one hip to the other).  They are instructed to do this to the beat of the metronome app set at 50 beats per minute.  This is done for 10 repetitions where 1 repetition is over and back. In Fig.~\ref{fig2}, we visualize a concussed and a non-concussed patient's data for left, right and cyclopean eye direction for visual motion sensitivity.
\end{itemize}

\subsection{Data splitting for Training and Testing}
\label{dst}
We acquired $1,373,333$ gaze data points from the VOMS and divided them into an 80/20 split of train and test samples. Each of samples have fourteen different features. Out of $1,098,666$ training samples, there are $1,089,990$ non-concussed and $8,676$ concussed data points.  And in the $274,667$ test samples, there are $272,498$ non-concussed and $2,169$ concussed data points. Next, we used a validation split of 0.1 on the training samples to validate the model. Following that, we use categorical class-weighting to train our models to address the class imbalance problem. It should be noted that Random Forest, Decision Tree, SVC, Logistic Regression, and Perceptron support class weighting during training.

Contrarily, we use a smaller training and test set for those methods that do not support class-weighting. The training set consists of 16,000 samples with 8,000 concussed and 8,000 non-concussed data. And, 4,000 test samples with 2,000 concussed and 2,000 non-concussed data. The following methods don't support class-weighting: AdaBoost, Gaussian Process Classifier, and Naive Bayes.

\subsection{Qualitative Evaluation}
In this section, we will elaborate on the qualitative evaluation using the One-class SVM and Isolation forest method for novelty detection.

\subsubsection{One-class SVM}
\begin{figure}[htp]
    \centering
    \includegraphics[width=0.9\textwidth]{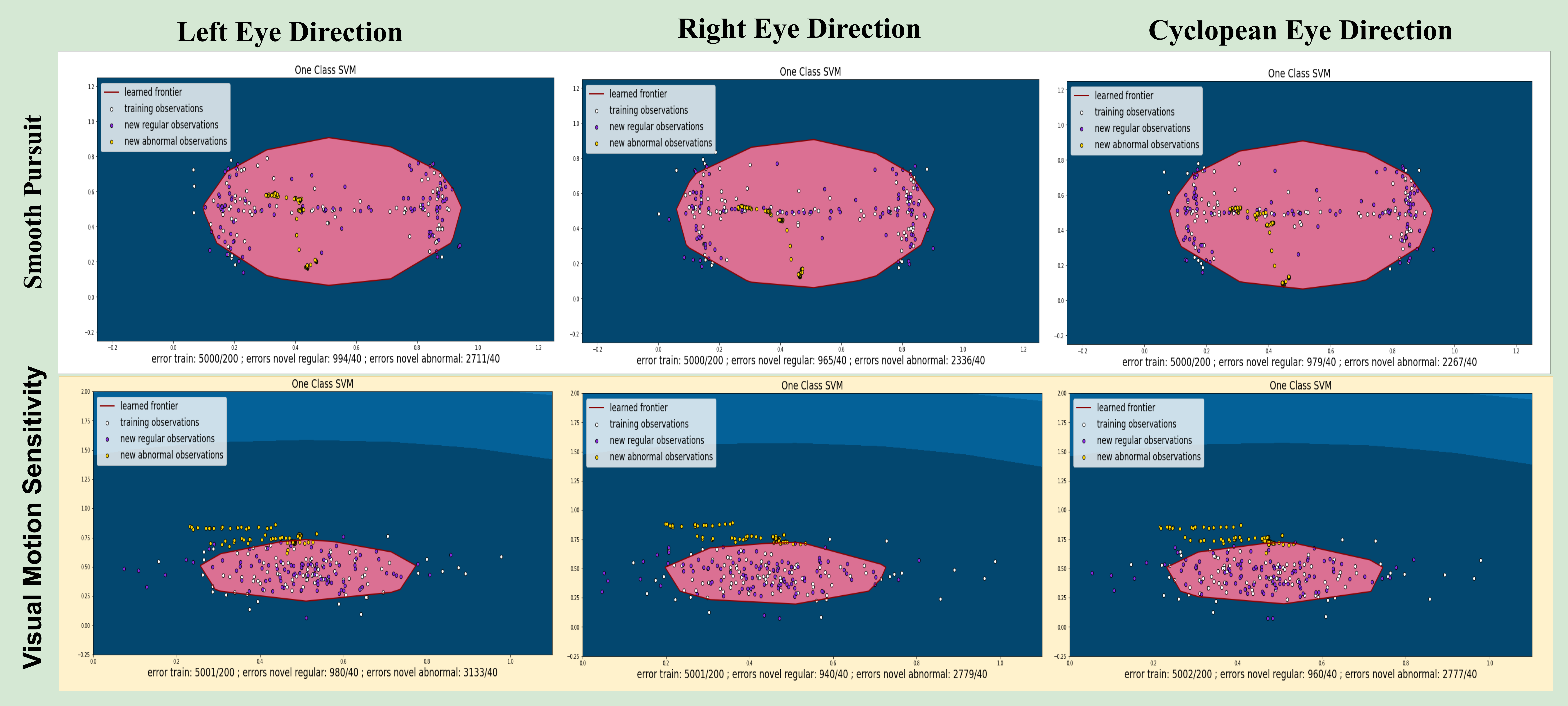}
    \caption{\textbf{One Class SVM:} The column I-II represents data points for Smooth Pursuit and Visual Motion Sensitivity. The row I-III represents Left eye, Right eye and Cyclopean eye directions. The white dots are training observations, purple dots are regular test observations and yellow dots are abnormal test observations }
    \label{fig3}
\end{figure}

In Fig.~\ref{fig3} we illustrate one-class SVM for both Smooth Pursuit and Visual Motion Sensitivity. Each Row of the figure visualizes Left eye, Right eye and Cyclopean eye direction data points. As training on millions of data point is computationally expensive, we use 10,000 training samples and 10,000 test samples for both Smooth Pursuit and Visual Motion Sensitivity. As we are doing novelty detection, we the training samples are all non-concussed control data points. However, the test consists of 5,000 concussed and 5,000 non-concussed data points. From the figure, it is apparent that the red line symbolizes learned boundary where the model is confident if the data point is novel or not (inside pink area). The white dots are training samples, violet dots are regular (non-concussed) test samples and yellow dots are abnormal (concussed) test samples. For visual motion sensitivity (column 2) the One-class SVM can easily isolate the outliers (concussed data points), however for Smooth Pursuit the it is quite challenging for detecting the outliers. Nonetheless, the outliers in Smooth pursuit are not concentrated in the two spectrum of the sphere, rather concentrated on the middle and bottom of the sphere. 

\subsubsection{Isolation Forest}
\begin{figure}[!tp]
    \centering
    \includegraphics[width=0.9\textwidth]{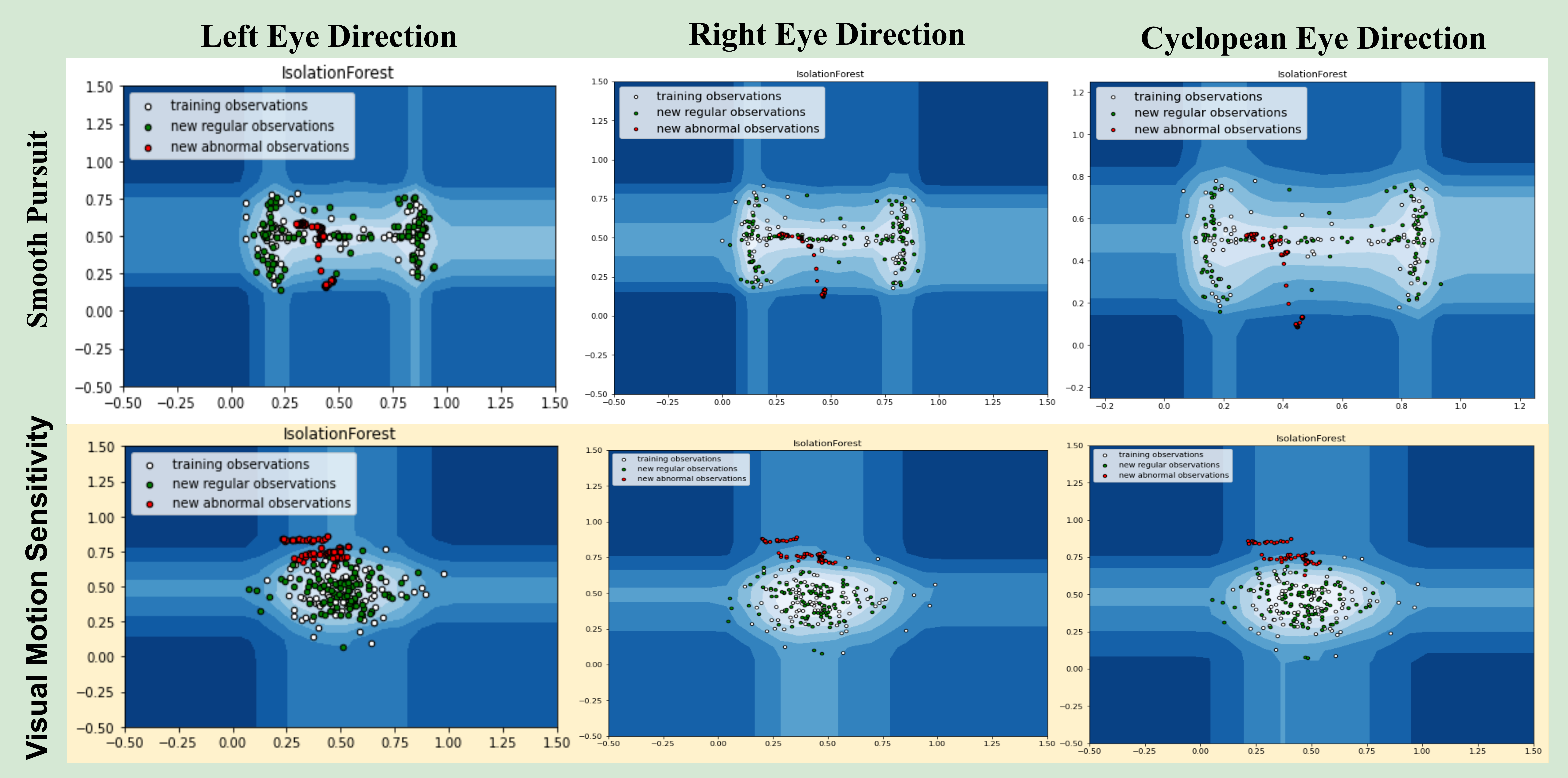}
    \caption{\textbf{Isolation Forest:} The column I-II represents data points for Smooth Pursuit and Visual Motion Sensitivity. The row I-III represents Left eye, Right eye and Cyclopean eye directions. The white dots are training observations, green dots are regular test observations and red dots are abnormal test observations }
    \label{fig4}
\end{figure}

In Fig.~\ref{fig4}, we illustrate the Isolation Forest for Smooth Pursuit and Visual Motion Sensitivity. Each Row of the figure visualizes the Left eye, Right eye, and Cyclopean eye direction data points. Similar to One-class SVM, we use 10,000 training and 10,000 test samples for Smooth Pursuit and Visual Motion Sensitivity. The test sample consisted of 5,000 concussed and 5,000 non-concussed data points. In the figure, white dots are trained observations (non-concussed), green dots are regular test observations (non-concussed), and red dots are abnormal test observations (concussed). For Visual Motion Sensitivity (column 1), the Isolation Forest can accurately distinguish the outliers (concussed data points) by finding the distance from the terminal to the root node of the feature space as discussed in \ref{iso}. Similarly, most outliers are isolated for the Smooth Pursuit data points (column 1). However, a few samples are still miscategorized as part of the regular observation in the figure.

\begin{table}[htp]
\caption{Evaluation on Test samples of \textbf{Smooth Pursuit} Dataset}
\centering
\begin{adjustbox}{width=1\textwidth}

\begin{tabular}{c|c|c|c|c|c|c|c|c}
\hline
Metrics     & Random Forest & AdaBoost & \begin{tabular}[c]{@{}c@{}}Gaussian Process\\  Classifier\end{tabular} & Decision Tree & Naive Bayes & SVM & \begin{tabular}[c]{@{}c@{}}Logistic\\ Regression\end{tabular} & Perceptron \\ \hline
Accuracy    & 100           & 100      & 100                                                                    & 100           & 98.4        & 100 & 99.4                                                          & 97.6       \\ \hline
Sensitivity & 100           & 100      & 100                                                                    & 100           & 96.8        & 100 & 99.4                                                          & 97.6       \\ \hline
Specificity & 100           & 100      & 100                                                                    & 100           & 100         & 100 & 99.5                                                          & 100        \\ \hline
Precision   & 100           & 100      & 100                                                                    & 100           & 98.4        & 100 & 80                                                            & 99.4       \\ \hline
F1-score    & 100           & 100      & 100                                                                    & 100           & 98.4        & 100 & 87.3                                                          & 98.3       \\ \hline
AUC         & 100           & 100      & 100                                                                    & 100           & 98.4        & 100 & 99.5                                                          & 98.8       \\ \hline
\end{tabular}
\end{adjustbox}
\label{table1}
\end{table}

\begin{table}[htp]
\caption{Evaluation on Test samples of \textbf{Visual Motion Sensitivity} Dataset}
\centering
\begin{adjustbox}{width=1\textwidth}

\begin{tabular}{c|c|c|c|c|c|c|c|c}
\hline
Metrics     & Random Forest & AdaBoost & \begin{tabular}[c]{@{}c@{}}Gaussian Process\\  Classifier\end{tabular} & Decision Tree & Naive Bayes & SVM  & \begin{tabular}[c]{@{}c@{}}Logistic\\ Regression\end{tabular} & Perceptron \\ \hline
Accuracy    & 99.9          & 99.9     & 99.9                                                                   & 99.9          & 96.5        & 99.9 & 96.2                                                          & 96.0       \\ \hline
Sensitivity & 100           & 99.8     & 99.9                                                                   & 100           & 93.5        & 99.9 & 96.1                                                          & 97.2       \\ \hline
Specificity & 99.9          & 100      & 99.9                                                                   & 99.9          & 99.6        & 99.9 & 96.9                                                          & 94.8       \\ \hline
Precision   & 99.9          & 99.9     & 99.9                                                                   & 99.9          & 96.7        & 99.9 & 98.2                                                          & 98.4       \\ \hline
F1-score    & 99.9          & 99.9     & 99.9                                                                   & 99.9          & 96.5        & 99.9 & 96.2                                                          & 97.5       \\ \hline
AUC         & 99.9          & 99.9     & 99.9                                                                   & 99.9          & 96.5        & 99.9 & 96.5                                                          & 96.0       \\ \hline
\end{tabular}
\end{adjustbox}
\label{table2}
\end{table}

\subsection{Quantitative Evaluation}
Using six standard metrics, we benchmarked eight trained models for quantitative evaluation on the test set. The metrics used are: i) Accuracy, ii) Sensitivity, iii) Specificity, iv) Precision, v) F1-score, and vi) AUC score. For the first experiment, we test on the Smooth Pursuit dataset, and the results are given in Table.~\ref{table1}. The table shows that Random Forest, AdaBoost, Gaussian Process Classifier, Decision Tree, and SVM achieved 100\% on all six metrics. Given that the Adaboost and Gaussian Process Classifier were trained on a smaller subset of the training samples, they achieved excellent scores compared to Logistic Regression and Perceptron. On the other hand, Naive Bayes being trained on a smaller subset achieved 98.4\% accuracy, precision, and F1-score, which is remarkable. It should be noted that we tested on all the samples. Moreover, the smaller subset of equal class-specific samples was chosen for training for that method which did not have built-in class-weighting parameters to prioritize minority classes (elaborated in section \ref{dst}).

We test on the Visual Motion Sensitivity dataset for the second experiment, and the results are given in Table.~\ref{table2}. From the table, we can see that Random Forest and Decision Tree have achieved 100\% sensitivity. Other than Naive Bayes, Logistic Regression, and Perceptron, all the other five methods reached 99.9\% accuracy, specificty, precision, f1-score, and AUC score. The trend for Gaussian Process Classifier and AdaBoost is similar to Smooth Pursuit, as they have reached 99.9\% in all metrics with a small number of training samples. Consqeuently, both evaluations validate that these methods accurately detect concussed data points.

\section{Conclusion}

In this paper, we experimented on the eye movement data of Smooth Pursuit and Visual Motion Sensitivity, which is generated by the virtual reality (VR) based VOMS assessment on detecting concussion. Again, we validated our results by comparing them with manual VOMS scores (0-10) to check the reliability of our experiment. We illustrate with multiple machine learning models that we can reliably and accurately classify concussed vs. non-concussed data using both Smooth Pursuit and Visual Motion Sensitivity information. One limitation of our study is we could not validate the other methods like saccades and VOR for concussion detection due to time-constraint. In the future, our target is to build a deep-learning architecture that can be trained on additional vestibular and ocular behavior analysis data points to demonstrate a VR-based standardized concussion detection method for athletic trainers and clinicians.

\bibliographystyle{ieeetr}
%
\bibliography{reference}

\end{document}